# A Robotic "Social Media" Controlled Observatory for Education and Research

David Lane[1]

[1]Burke-Gaffney Observatory, Saint Mary's University, Halifax, Canada, dlane@ap.smu.ca

**Abstract:** I describe the world's first robotic observatory to interact with its observers entirely using the social media platforms Facebook or Twitter. The telescope "tweets" what it's doing, posts live images, and responds to observer commands through a comprehensive command set. Observation requests are queued and observed by a responsive queue engine. Its architecture, social media based image processing capability and several usage examples are also described.

**Keywords**: social media; robotic telescope; education; research

**Introduction**

In 2015, the Burke-Gaffney Observatory (BGO) in Nova Scotia became the world's first observatory to possess a fully-automatic Twitter interface (Capern 2016). Since then, about 4,100 observation requests have been fulfilled for university and high school astronomy students, beginner and advanced amateur astronomers, a few research-focused observers, and for AAVSO-organized observation campaigns.

A second observatory — the Abbey Ridge Observatory (Lane, 2017a) owned by the author — has also gone live.

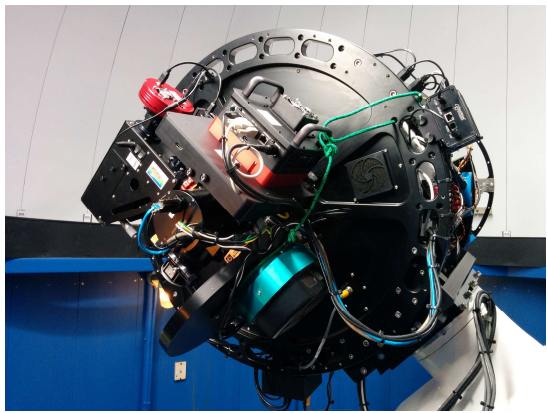

*Figure 1  The BGO's Planewave CDK24 telescope and its instruments: Optec Perseus 4-port selector with Apogee Aspen CG16M CCD (10 filters – luminance, red, green blue, B, V, R, I, OIII and H-alpha), SBIG STLX11002 CCD (8 filters), Shelyak LHIRES II spectrograph (manual use only), and 2-inch eyepiece port. Software integration of the second (SBIG) camera is in process. The telescope is mounted on the observatory's original Ealing mount upgraded with a Sidereal Technology Servo II controller.*

As part of a major equipment upgrade to the 45-year-old observatory, it was automated and an easy to use interface has been developed that makes the observatory accessible to everyone — young and old — using the social media tools they were already using. The primary motivation was to better reach young people and the "screen" generation.

The Twitter social media platform was chosen as the initial interface — it had the advantage of being commonly used by many university students and is accessible from any web browser and all tablets and smart phones using the Twitter app. Later, Facebook Messenger and e-mail interfaces were added. The e-mail interface was developed mainly for our own students, so they would not be required to use a proprietary social media service.

**How it Works**

Observers interact with the observatory by sending it messages ("commands") in a particular format which it replies to, usually within a minute. Twitter calls these messages "tweets" when publicly viewable and "direct messages" when private. The Facebook interface uses its private Facebook Messenger service. It also communicates what it's doing as Twitter "tweets" and posts to its Facebook page.

About twenty hashtag commands (words with a # prefix) are understood that allow an observer to make imaging requests, manage their requests, and obtain the observatory's status and weather information.

An associated website contains searchable listings of all requested and completed observations, comprehensive reference information, FAQs, and much more.

Figure 2 shows an example "conversation" using Twitter direct messages.

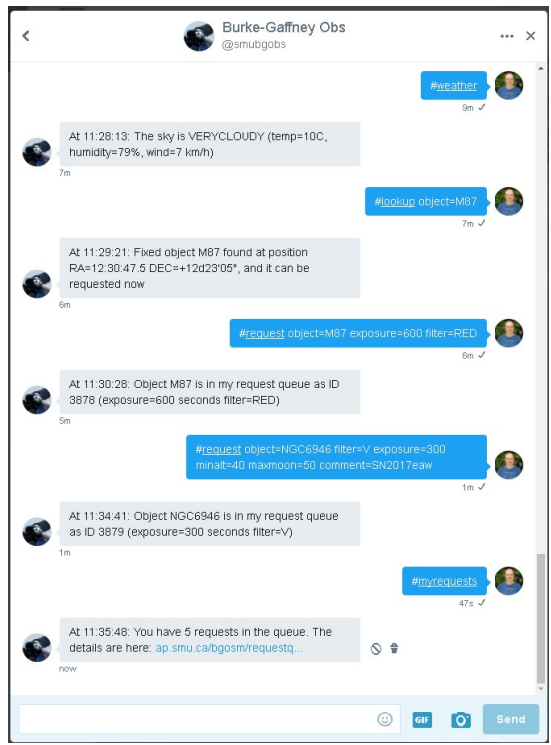

*Figure 2   An example conversation between an observer and the observatory. Commands are hashtags (words prefixed by a # character) and command options use the syntax 'parameter=value'.*

While the observatory listens for messages 24/7, of course it can only do its imaging work at night, so it queues imaging "requests" and runs them from the queue on clear nights.

Before requests are accepted, they are validated. Only objects in the telescope's object database and those reasonably-observable within the next month are accepted. The object database originates from queries made to The Earth Centered Universe Planetarium software (Lane 2016) — it includes most common fixed objects and currently observable asteroids and comets — the observer can add objects to the database too.

Several types of observations are supported including: single filter, multi-filter, and time series (single or multi-filter). Each observer is granted "permissions" that, for example, determine the permitted exposure types, set limits on exposure time, or the number of requests in the queue. Observers can specify the filters used, exposure times, minimum altitude, maximum moon illumination, position and focus offsets, etc. Whatever not specified will default to something reasonable — it even auto-selects the exposure and filter automatically for the Moon and bright planets and based on the type of deep sky object. While these settings provide flexibility for experts, beginners can request images of common targets with a message as simple as:

#request object=M57

Accepted requests are assigned a unique request ID number.

Each night after it becomes somewhat dark, the observatory uses its cloud sensor to monitor the sky. If the skies are clear, it powers, opens the dome, and focuses the camera and syncs the telescope's position. If it wasn't clear when it became dark, it waits for it to clear.

After the sky has become fully dark and until shortly after the beginning of morning twilight the observatory works through its queued observation requests, pausing if it becomes cloudy and restarting when it clears. When an observation is completed, the observer is notified. For single-filter observations, "quick look" images are posted within a few seconds!

After the dome as closed at dawn, whatever calibration images are needed at taken. The night's images are calibrated and combined, then they are Robotic Telescopes, Student Research and Education Conference, June 2017posted to the "completed" queue on the website. The observatory then notifies each observer individually that they have images ready. The cycle repeats for the next night.

Observers are provided instrumentally calibrated FITS images and additionally, for single-filter observations, automatically-stretched JPEG images. All images produced are publicly available.

While not presently accessible from the social media interface, aperture photometry can also be done for specified stars, including the automatic production of AAVSO photometry files.

**System Architecture**

The observatory's software operates on current versions of Microsoft Windows. It uses the ASCOM Standards (Denny, 2017a) hardware drivers, MaxIm DL (George, 2017) for image acquisition and processing, and PinPoint (Denny, 2017b) for plate solving. It was designed with portability in mind — as such, it is quite adaptable for use at other observatories with hardware compatible with ASCOM and MaxIm DL. Figure 3 shows a block diagram of the system.

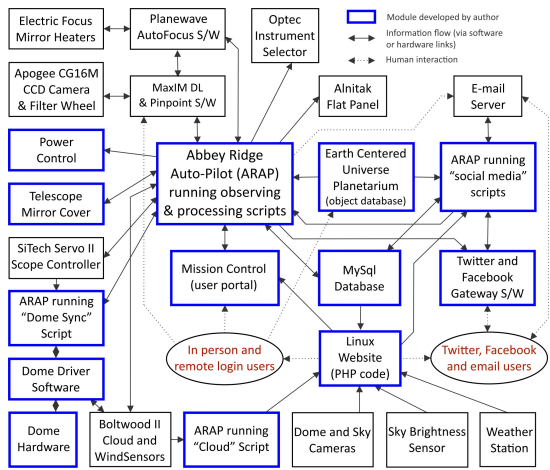

*Figure 3*  A block diagram showing the observatory's major hardware and software components and their relationships. All blocks with thicker boxes were designed and developed by the author.

The observatory's software is written mostly in scripts that are run in the Abbey Ridge Auto-Pilot (ARAP) software (Lane, 2017b), which is an Observatory Automation Language originally developed by the author to automate the operations and image processing at the Abbey Ridge Observatory. It has been under constant development since 2003 and has been significantly enhanced for this project.

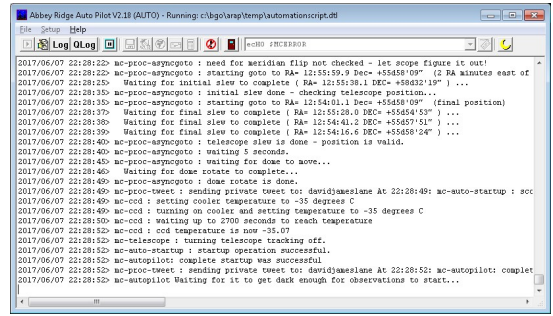

*Figure 4*  The run screen of the Abbey Ridge Auto-Pilot (ARAP) software — several copies of ARAP run scripts to implement the observatory's software subsystems (observing, social media, dome syncing, website weather graphs, etc.).

ARAP runs text-based scripts containing English-like commands, in effect creating a rich observatory control language, that makes it very easy to automate observatory functions. ARAP itself and the other applications developed (Earth Centered Universe planetarium, Mission Control user portal, and the Twitter and Facebook Gateway) are written in Embarcadero Delphi (Object Pascal). The website database pages are written in PHP. In all, there are about 100,000 lines of program code that make it all work!

**Notable Features**

*Responsive Queue*

Many robotic telescopes plan observations nightly based on the known variables (what's in the queue, what is visible, the length of night, observing requirements, sorting by efficiency, priorities, etc.).

In contrast, the BGO's queue was developed to be more responsive to the expected mode of operation in a social media environment. For example, observers with different priorities (eg. our students with project deadlines, researchers, astronomy enthusiasts, etc.), rather poor weather at our sea-level and seaside location, different types of observations (single filter, multi-filter, and time series), targets of opportunity that need to run urgently (eg. supernova confirmation or near Earth asteroid), and observations requiring specific conditions (minimum altitude, maximum moon illumination, or observations that need to take place during a specific time window (eg. eclipsing binaries).

Each time the telescope is ready to run an observation, it scans through the observation queue to see which requests can be observed right now. The factors it checks include:

- that the object's altitude is above the minimum at the beginning and end of the observation;
- that the end of the observation will complete before morning twilight begins;
- that the Moon is not too bright and is not too nearby in the sky.

If no observations can be run at the moment, the observatory posts a message welcoming live observations.

From the list of observable requests, a "points game" is played and the request with the highest score is run! The factors scored include:

- length of time in the queue - older requests are favoured over more recent requests.
- object visibility - objects that can be observed for a shorter period of time that night than others are favoured.
- observer priority - each observer is assigned a priority, either more or less than the default.
- group priority - each observer is a member of a group of observers and each group is assigned a priority, either more or less than the default. This is used to give groups of observers (eg. our students and those working on special projects) greater priority.
- observation priority - each observation request can be given a priority, either more or less than the default.
- same or opposite side of pier - our telescope mount is a German Equatorial, meaning it has to "flip" when moving from the east-to-west or west-to-east sides of the sky. This takes a few minutes so a penalty is applied if it is needed.
- distance from current telescope position - objects near where the telescope is currently pointed are favoured.

The setting of one or more of the "priority" values (observer, group, and observation) is the principal way to ensure that observations run when they need to. For example, students in university classes are assigned to a group and the group is given an elevated priority greater than the "default" observers. For targets of opportunity or observations that need to run at a particular time based on an ephemeris (eg. an exoplanet transit), one or more priority values are raised high enough so to ensure they run as soon as they can be given other provided conditions. Presently, however an observation already underway cannot be interrupted. Likewise, observations given a very low priority would only run when the telescope is idle (when no other requests were observable).

*Automatically Producing Good Quality JPEG Images*

While FITS files are important for quantitative measurements and post-processing, automatically producing good 8-bit images suitable for posting to social media is necessary so the observers can see results quickly and without special software.

Given the large dynamic range and the wide variety of astronomical sources, this is not easy to realize!

However, after some research and experimentation, an algorithm that uses the ArcSinH function was selected to stretch images (Lupton et al, 2003). This algorithm works remarkably well for a wide variety of object types. See Figure 5 for a typical result.

A histogram of the FITS image is computed to determine the maximum and median ADU values. The background level is set to the median, which for most images equates to just above the sky background level. Each pixel is then stretched as follows:

*StretchedPixel=*
*ArcSinH(Pixel×Scale×Power)×Damper*

where:

*Damper=(256-Base)/*
*(ArcSinH((256-Base)×Power))*

*Scale=(256-Base)/(Maximum-Median)*

After some experimentation, settings of Base=15 and Power=5 work well for most objects.

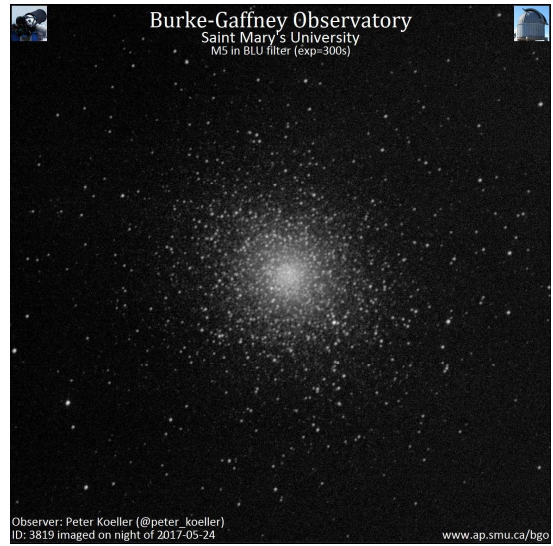

*Figure 5  This image of globular cluster M5 is an example of an automatically-processed and ArcSinH-stretched jpeg image as would have been tweeted to the observer and posted to the website image archive. Images are annotated with text that detail observation information and personally identifies the observer.*

### Image Processing via Social Media

Soon after launch, it became apparent that most novice observers, which would include first year astronomy and high school students, do not have access to the software needed to read and process FITS files. While the automatically-produced JPEG images described above were good, many observers wanted to improve them but couldn't easily do so.

Even the clever ones, after importing the FITS files into non-astronomical image processing software, quickly realized that they could not even easily stack or make colour images because the stars did not line up!

As a result, a set of image processing operations was developed that can be easily done entirely from social media commands. The input image to a processing operation can be either an original image or the output from an earlier processing operation.

For example, to stack 3 images by averaging is as simple as sending the message:

*#process id=1234,1235,1254 step=stack stack=average*

This aligns the input images on their stars and stacks them to form the output image. It then sends the observer a web link to the result (with the FITS image). If the telescope is not presently observing, the operation is usually done within a minute, otherwise it is queued. A more advanced example is described in Figure 6.

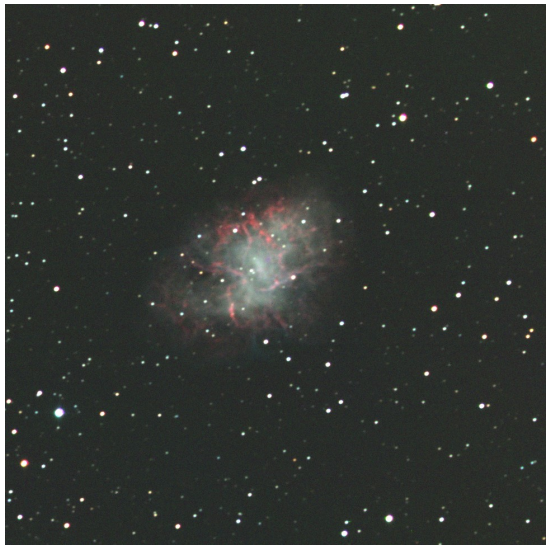

*Figure 6   This image of the Crab Nebula (M1) was processed using the social media interface. It involved the following steps: stacking two 5-minute luminance frames, combining stacked luminance and 5-minute red, green, and blue filtered images into a colour image, cropping the image edges, removing the background gradient, adjusting the colour saturation, image sharpening with an FFT filter, and finally a gamma stretch.*

Image processing operations presently supported include: flip, mirror, flatten background, background gradient removal, rotate, crop, resize, FFT filtering (high and low pass), kernel filtering (high pass, low pass, hot pixel, cold pixel, etc.), image stacking, colour combinations (RGB and LRGB), colour saturation, colour balance, and stretches (linear, log, and gamma).

**Examples of its Usage**

*University Courses*

The social media interface has been used extensively at the BGO's host university over the past two years where we run a four-year BSc Astro-Physics program. First year students typically request images and interpret and comment on the images qualitatively. Second to fourth year students have, for example, constructed H-R diagrams, observed exoplanet transits, created RR Lyrae star light curves, or measured positions of asteroids and computed an orbit. Figure 7 shows the data obtained for one such project.

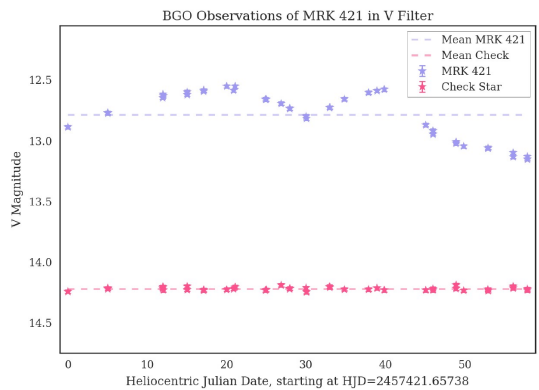

*Figure 7   About two months of V-band photometry taken by 3rd year Astro-Physics student Tiffany Fields for an Observational Astronomy course. This was part of an analysis of the V and B-band variability of active galaxy MRK 421. Prior to automation and an easy to use interface, it would have been impossible to obtain so much data within the time constraints of a university semester.*

*AAVSO Campaign Observing*

The author has participated in many AAVSO observing campaigns, an example of which is shown in Figure 8.

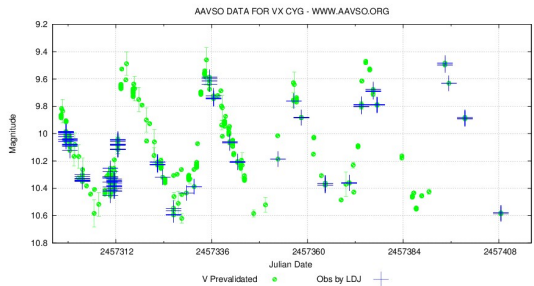

*Figure 8   AAVSO International Database V-band photometry of VX CYG from the summer of 2016 as part of a Cepheid monitoring campaign. The observations indicated by the blue crosses were taken by the BGO's social media interface. The photometry was also automatically measured.*

### High School Astronomy Classes

We have a strong interest in having the observatory be used by High School level students, particularly in the spring when our university classes are not running.

Here are comments from two high school teachers who have used it with their students:

> *I am a high school science teacher at Maples Collegiate High School in Winnipeg, Manitoba. Over the past few years, my students and I have used remote telescopes in Siding Springs Australia to take images of the night sky. This year in my Astronomy 31G course, I developed a project that involved taking images of the Northern Hemisphere skies, using the Burke Gaffney Observatory (BGO) at Saint Mary's University. It was a unique project in that students had to use their Twitter or Facebook accounts to communicate with the remote BGO. The students first had to research what objects they wanted to take an image of, as well as familiarize themselves with the Northern Hemisphere sky using the program called Stellarium. The students had to work out when their object was up in the night sky for viewing and send a request for their object ahead of time. Many of the students who sent in requests got back amazing images. The remote BGO was easy to use and gave my students the opportunity to explore the night sky in a way few get a chance to do!*
>
> *- Andrea Misner*

> *The students from the distance-learning courses in New Brunswick had the opportunity to use the BGO (Burke-Gaffney Observatory) on a few occasions. The goal was to help students discover the main astronomical bodies around Earth (eg. galaxies, star clusters) with a different approach. After familiarizing themselves with the Messier catalogue, the students had to choose, according to their own preferences, different celestial bodies and send a request to the BGO to obtain a photograph. After obtaining the photographs, students had to describe in a project, for example, the differences in morphology, as well as the origin and evolution of different galaxy types. Being able to have different settings options (ex. filter, exposure time) and to obtain real pictures made it really interesting for the students. Thanks to Saint Mary's University to have given us this opportunity.*
>
> *- Kevin Burke*

### Future Development

The capabilities described in this paper are fully implemented and reliable, however the development of new capabilities and improvements to the existing functionality are ongoing.

Some features under-development or being considered include:

- Multi-camera and multi-telescope support
- Automatic auto-guiding
- Dithering during image acquisition

- Make photometry available from social media interface

- Web-interface

- Reserved "live" observing windows

**Acknowledgements**

This project would not have been possible without a substantial donation from philanthropist Dr. Ralph Medjuck, who believed in our vision to make the BGO much more accessible to our astrophysics students, high school students, and others. Dr. Medjuck also provided funding to build the observatory in 1972.

I also thank Dr. Adam Sarty, my Associate Dean of Science (Outreach), for funding the travel to this conference.

**References**

Capern, Rory, 2016, 10 amazing Canadian stories on Twitter, Date of access,12/06/2017. http://blog.twitter.com/official/en_ca/a/2016/10-amazing-canadian-stories-on-twitter.html

Denny, Bob, 2017a, ASCOM Standards, Date of access: 13/06/2017. http://www.ascom-standards.org

Denny, Bob, 2017b, PinPoint software, Date of access: 13/06/2017. http://pinpoint.dc3.com

George, Doug et al, MaxIm DL software, Date of access 13/06/2017. http://diffractionlimited.com/product/maxim-dl

Lane, David, 2016, Earth Centered Universe Planetarium software, Date of access: 2017/11/02. http://nova-astro.com/ecupro.html

Lane, David, 2017a, Abbey Ridge Observatory, Date of access: 12/06/2017. http://www.abbeyridgeobservatory.ca

Lane, David, 2017b, Abbey Ridge Auto-Pilot, Date of access: 13/06/2017. http://www.abbeyridgeobservatory.ca/about/abbey-ridge-auto-pilot

Lupton, Robert et al, 2003, Preparing Red-Green-Blue (RGB) Images from CCD Data. PASP, 116, 133-1.